# Rotation effect in optical imaging of symmetrical figures in an opaque screen

A. Bekirov,[*,†] V Sitnyansky[†], S. Senotrusova[†], B. Luk'yanchuk, I. Yaminsky and A. Fedyanin

*Faculty of Physics, Lomonosov Moscow State University, Moscow 119991, Russia*
[†]*The authors contributed equally to this Letter.*
[*] *bekirovar@my.msu.ru*

**A novel effect of rotation in the optical image of structures in an opaque screen is presented. Theoretical calculations and experimental validation are presented. The features and the conditions of this effect are discussed.**

**Introduction**. According to geometrical optics, an optical image of an object can be observed when it is positioned at the focal point of a lens. When the object is moved above or below this focal point, the image loses clarity and becomes blurred. This blurring occurs because the rays emanating from a point source do not converge to a single point in the observation plane. Geometrical optics posits that the position at which the image of an object can be observed is unique. However, light also exhibits wave properties, which prevent it from being focused to a single point due to diffraction. Consequently, it is more accurate to describe the field distribution at the lens focus using wave optics.

In this context, new phenomena may emerge that are not accounted for within the framework of geometrical optics. This paper explores a novel effect: the rotation of images of planar figures when the sample is displaced from the lens focus. Specifically, in addition to the image formed in the focal plane, as described by geometrical optics, a rotated image of the original object can be observed both above and below the focal position. It is essential that the radiation emitted from the observed sample is coherent for this effect to manifest. This phenomenon can be observed for specific types of planar figures whose dimensions are comparable to the width of the point spread function (PSF).

**Theoretical treatment.** According to Abbe's theory, the optical image created by a lens in the focal plane can be derived using the inverse Fourier transform of the source field. This approach involves discarding the evanescent components of the spatial spectrum, which do not contribute to the image formation in the focal plane. The mathematical representation of this process can be expressed as follows [1, 2]:

$$\mathbf{E}^{im}(x,y,z) = \iint_{k_x^2+k_y^2 \leq k^2} \tilde{\mathbf{E}}^*(k_x,k_y,z_o) e^{-i[k_x x + k_y y + k_z(z-z_o)]} \frac{dk_x dk_y}{(2\pi k)^2}. \quad (1)$$

Here, $\tilde{\mathbf{E}}$ is the two-dimensional Fourier transform of the source field $\mathbf{E}$ in the xy plane for $z = z_o$, $k=2\pi/\lambda$, $k_z^2 = k^2 - k_x^2 - k_y^2$, $\lambda$ is wavelength of light. Formula (1) can be rewritten as a diffraction integral in the forms:

$$\mathbf{E}^{im}(\mathbf{r}) = \frac{-k^2}{4\pi} \iint_\Gamma \left( \frac{-i\mu}{n} [\mathbf{n}, \tilde{\mathbf{H}}^*] G + \left[ [\mathbf{n}, \tilde{\mathbf{H}}^*], \nabla G \right] + (\mathbf{n}, \tilde{\mathbf{E}}^*) \nabla G \right) dS \quad (2)$$

here $G = \exp(ik|r-r_0|)/k|r-r_0|$, $\nabla = k^{-1}\partial/\partial \mathbf{r}_0$, $\Gamma$ is surface homotopic to the plane. Formulas (1-2) model the optical image created by an aberration-free system with 1:1 magnification and an infinite aperture. In the case of a point source field, the field of which is a solution to the Helmoltz equation:

$$\Delta \mathbf{E} + k^2 \mathbf{E} = -4\pi k^{-1} \delta(\mathbf{r} - \mathbf{r}_1) \mathbf{e}_x, \quad (3)$$

in the plane of the source $z = z_1$, the image field is expressed by the formula [3]:

$$\mathbf{E}^{im}(\mathbf{r}) = -i \sin(k\rho)/(k\rho) \mathbf{e}_x, \quad (4)$$

here $\rho = \sqrt{(x-x_1)^2 + (y-y_1)^2}$. Let us consider two coherent sources of type (4), located along the x-axis, at distances of wavelength λ. In this case, two maxima in the image field (3) distribution in the plane of the sources can be observed. For planes $z = z_1 \pm \lambda$, due to interference, two maxima merge into one, located between the sources and elongated along the y-axis. Talbot effect [4] seems like to this one, it occurs for light diffraction on periodic structures, but the alternation of maxima occurs in a more complex manner and is observed in a diffracted field, not in the image field. The shift of the maximum in the image for two sources leads to the appearance of a mirror image of flat objects when the observation plane is shifted. Let us consider the image field (1-3) obtained from the diffraction of a normally propagating plane wave on a 100 nm thick gold screen with an aperture in the form of a triangle and a six-pointed star. Diffraction calculations were performed using numerical simulations based on the finite difference time domain method implemented in the Ansys Lumerical FDTD application package. This method approximates Maxwell's equations using finite differences in the time and space domains. The image field was calculated according to formula (3). The results are presented in Fig. 2, you can observe both direct and mirrored images in planes $z = z_1$ and $z = z_1 \pm \lambda$ accordingly, here $z_1$ is a plane of a screen. This effect can be explained by interference between the sources at the vertices of the figure, similar to what was considered for two sources. When viewing objects much larger than the wavelength or in incoherent light, the effect is not observed.

geometric optics), both above and below the image formation plane, mirrored images can be observed, which is in accordance with the theory presented above. Note that in the plane of focus the image is inverted, which is explained by the laws of geometric optics. In planes other than focal planes, the image is rotated again.

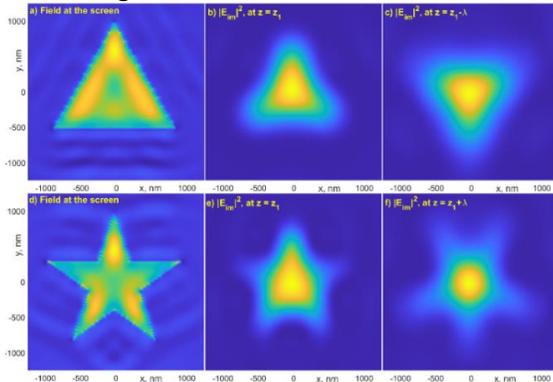

Fig. 1. (a), (d) - real field in the plane of the screen. Theoretical calculation of the image (3) obtained from the diffraction of a plane wave on a metal screen with a hole in the form of a triangle (b-c) and a star (e-f). The image is calculated in planes $z = z_1$, $z = z_1 \pm \lambda$.

Let us note in more detail the conditions under which this effect is observed in the presented theory. Firstly, the construction of the image must occur in coherent light, i.e. the radius of coherence of radiation incident on the screen must be of the order of the size of the structure. Secondly, the distance from the center to the edge of the structure must be comparable to the PSF size of the optical system. In the examples we have considered, the sizes of the PSF are of the order of $\lambda$, therefore, the structures themselves have the same size. Thirdly, the distribution of the electromagnetic field in the plane of the object should be close to uniform. The structures must have a certain symmetry, in which the interference of the image from the center and edges of the figure leads to inversion in planes other than the focal one. It should be noted that the mirrored images is observed in a limited range of planes near the focus, see Fig. 1. A similar effect was noted when calculating the virtual image in a spherical particle [5]. However, due to the reasons stated above, the effect was not observed in works considering such structures [6, 7].

We have considered the image of the created idealized optical system, described by formulas (1-3). In practice, the image is constructed using lenses. Does this effect occur in the case of real lenses? We designed a Fresnel lens placed on a gold screen with a triangle or star shaped hole in it. The position of the lens was determined in such a way that the screen with holes was located in its focal plane. The system forms an image in a transmission scheme. A plane wave with a wavelength of 1.2 μm is selected as the field source. The material chosen for the Fresnel lens is photopolymer ORMOSIL (SZ2080), the refractive index at this wavelength is $n=1.454+0.013i$ [8]. The fields were read using monitors located in the focal area, the results are presented in Fig. 3. In the area of the geometric focus of the lens, an inverted image of the object can be observed (which is consistent with

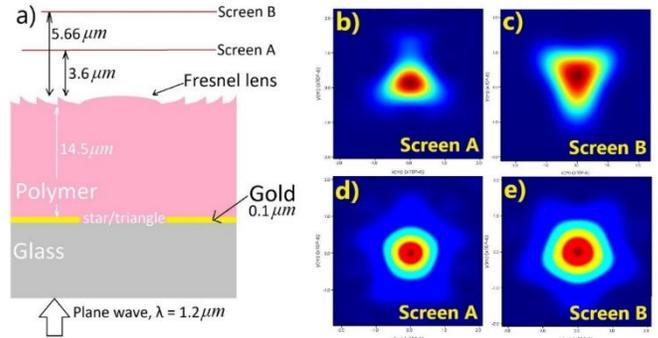

Fig. 2. (a) General view of the calculation geometry, a field with a wavelength of 1.2 mkm propagates through a gold screen with a hole in the form of a star or triangle. An image formed by a Fresnel lens for a triangle (b), (c) and a five-pointed star (d), (e) taken from screen A and B. For (d), (e), the contrast is enhanced. Due to blurring, only two images were observed for the star. The refractive index of the photopolymer is $n=1.454+0.013i$.

This calculation is a verification of the results obtained using formulas (1-3) since this lens on the structure creates an image with a magnification close to 1:1. The geometry we presented can serve as a basis for direct experimental verification of the effect. In our work we used a simpler approach and examined this effect using a microscope.

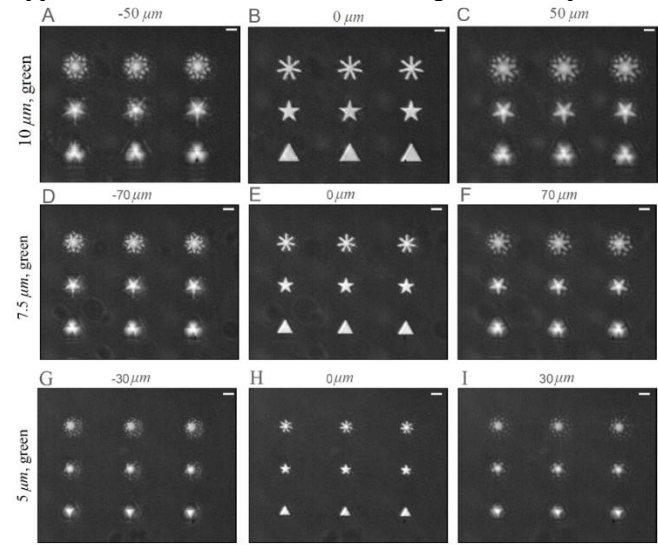

Fig. 3. Visualization of the effect of displaying figures horizontally at different focus positions. The focus position is indicated above each image, respectively, where 0 μm is the focus position on the sample surface. The size of the figures is respectively: (A-C) – 10 μm, (D-F) – 7.5 μm, (G-I) – 5 μm. The scale bar in the upper right corner of the image corresponds to 10 μm.

**Experiment.** To experimentally observe the effect we presented, we made a 80 nm thickness gold plate a located on a glass substrate with different shape structures on it, such as triangle, five and seven pointed star. The sizes of the samples were *5*, *7.5* and *10 μm* from central part to the vertex. See Supplemental 1 about detail of the sample .The experiment was carried out using a Nikon Ti-U Eclipse microscope (*NA=0.6*) with green LED illumination in transmitted light mode. The optical images at the different focus distances are represented in Fig. 4. The effect is most pronounced for the five-pointed star. Rotation for the triangular figure can only be observed for the *5 μm* sample. The effect was not observed on structures with dimensions of the order of the wavelength. This can be explained by the fact that this microscope constructs an image with significant optical magnification.

For a complete theoretical modeling of this effect in a microscope, it is necessary to find the point spread function exactly. This is not an easy task, but the results we have presented indicate the presence of this effect in real systems. Note that similar results were obtained for a blue LED source.

**Acknowledgements.** AB thanks support from Foundation for the Development of Theoretical Physics and Mathematics «BASIS».
**Disclosures**. The authors declare no conflicts of interest.
**Data availability.** Data underlying the results presented in this paper are not publicly available at this time but may be obtained from the authors upon reasonable request.